\documentclass[a4paper,11pt,headings=big]{scrartcl}
\usepackage{amsmath,amssymb}
\usepackage{bbm}
\usepackage{graphicx}
\usepackage{multirow}

\addtokomafont{disposition}{\rmfamily\boldmath}
\let\tmptitle\title\renewcommand{\title}[1]{\tmptitle{\LARGE #1}}
\let\tmpauthor\author\renewcommand{\author}[1]{\tmpauthor{\large #1}}
\let\tmpdate\date\renewcommand{\date}[1]{\tmpdate{\normalsize #1}}
\newcommand{\abstrct}[1]{\begin{abstract}\vspace{-2em}\small\noindent#1\end{abstract}}

\title{\Large
Beyond the standard MSSM
}
\date{}
\author{
Riccardo Barbieri
\\ \normalsize\itshape
Scuola Normale Superiore and INFN, Piazza dei Cavalieri 7, 56126 Pisa, Italy
}

\begin{document}

\maketitle

\abstrct{%
If supersymmetry is relevant at the Fermi scale, the lack of any direct signal so far may require going beyond the Minimal Supersymmetric Standard Model. In this talk I briefly summarize a simple and concrete extension of the MSSM that takes these issues, including a way to address the flavour and the CP problems. In its fully natural range of parameters, the expected signals for LHC, dark matter and flavour physics are clear and generally quite different from the ones of the MSSM. Gauge coupling unification may be only approximate.
}

\section{Introduction}
\label{sec:Intro}

The Fermi scale may either find its origin in some  weak coupling physics, like in the Standard Model (SM) itself with a not too heavy Higgs boson, or in some new strong force. The first case is currently preferred by indirect data, like the ElectroWeak Precision Tests (EWPT), or even by gauge coupling unification. Supersymmetry is the leading candidate for the weak coupling picture of ElectroWeak Symmetry Breaking (EWSB).

This last statement should however be contrasted with the lack of any direct signal so far. With reference with the expectations of the  Minimal Supersymmetric Standard Model (MSSM), these unseen signals are: i) no Higgs boson, ii) no s-particles, iii) no clear deviation from the SM in flavour and in CP physics. This may either mean that such direct signals are there, so to say  around the corner, just waiting to be discovered  or, on the contrary,  that weak-scale supersymmetry is not realized in nature. With LHC ready to produce data with a significant luminosity, it looks unreasonable to take now a strong position on this alternative. Rather I find more useful to give consideration to a third possibility:  Does a simple and concrete extension exist of the standard MSSM that can explain the lack of signals so far? This is not a new question. My  view on it is based on the following two points:
\begin{itemize}
\item The lack of Higgs boson and of s-particle signals may be related issues: if the Higgs boson(s) can be made heavier than in the standard MSSM, this generally relaxes in a corresponding way the naturalness bounds on all the s-particle masses;
\item The suppression of any new source of flavour and CP violation may have something to do with the heaviness of all the s-fermions not coupled via the top Yukawa coupling to the Higgs system, i.e. to the relative lightness of the two stops and the left-handed sbottom only.
\end{itemize}

\begin{figure}[tb]
\centering
\includegraphics[width=10cm]{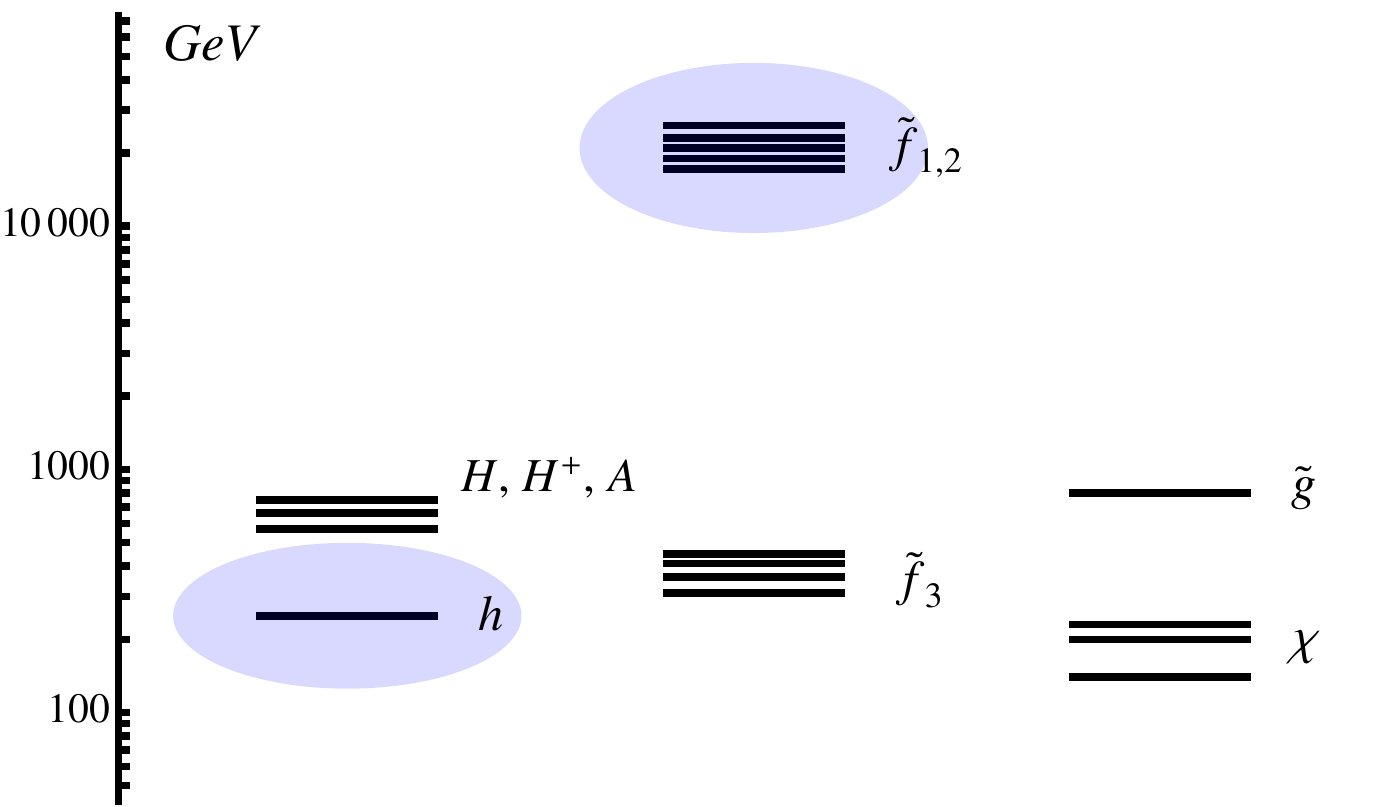} \caption{{\small A representative Non Standard Supersymmetric Spectrum with $m_h = 200\div 300$ GeV and $m_{\tilde{f}_{1,2}}\gtrsim 10$ TeV.}}
\label{spettro2}
\end{figure}

Although independently valid, naturalness  hints at a relation between these two points. I maintain that there is a simple extension of the MSSM that  realizes these points in an overall coherent way, with clear and generally quite different signatures from the ones of the standard MSSM. The key feature is summarized by the typical s-particle spectrum shown in Fig. 1, up to  details to be specified below \cite{Barbieri:2010pd}.

\section{Supersymmetry without a light Higgs boson}
\label{nolightHiggs}

Ways have been examined to evade the tree level upper bound on the mass of the lightest Higgs boson, valid in the MSSM, $m_h < m_Z \cos{2\beta}$. They rest on  a modification of the Higgs potential either though a D-term or though an F-term. The first requires an extension of the gauge group, necessarily also acting  on all the matter supermultiplets via a sizable new gauge coupling \cite{Batra:2003nj,Batra:2004vc,Maloney:2004rc}. The second may involve an extension of the Higgs system only, the simplest example being represented by the supersymmetric Yukawa coupling of the NMSSM, $\lambda \hat{S}\hat{H}_u\hat{H}_d$ \cite{Harnik:2003rs,Chang:2004db,Delgado:2005fq, Barbieri:2006bg}. As well known, in the last case the upper bound on the mass of lightest scalar gets modified to
\begin{equation}
m_h \leq m_Z 
(\cos^2{2\beta} + \frac{2\lambda^2}{g^2 + g^{\prime 2}}\sin^2{2\beta})^{1/2}.
\label{m_h}
\end{equation}
 If one wants   $m_h$ increased by a significant amount, say between 200 and 300 GeV, $\lambda$ cannot be too small,  between 1.5 and 2. On the other hand I do not find useful to consider raising the Higgs boson mass by an effective Lagrangian approach, since the new mass scale involved has to be close to the Fermi scale, which in turn obscures the weak coupling picture of EWSB and threatens the control of the EWPT.

To me the NMSSM with a largish $\lambda$, thus dubbed $\lambda$SUSY, emerges as the simplest  concrete possibility to raise significantly the Higgs boson mass. It can be shown that the upper bound on the Higgs mass set in the SM by the EWPT is automatically evaded in $\lambda$SUSY .Furthermore, as anticipated on general grounds, the naturalness bounds on the s-particle masses are indeed relaxed by a significant amount: with $\lambda = 2$ and a Higgs boson as heavy as 250 GeV the gluino can weight up to 1.5 TeV with less then 20\% fine tuning \cite{Barbieri:2006bg}. 

\begin{figure}[thb]
\begin{center}
\includegraphics[width=0.55\textwidth]{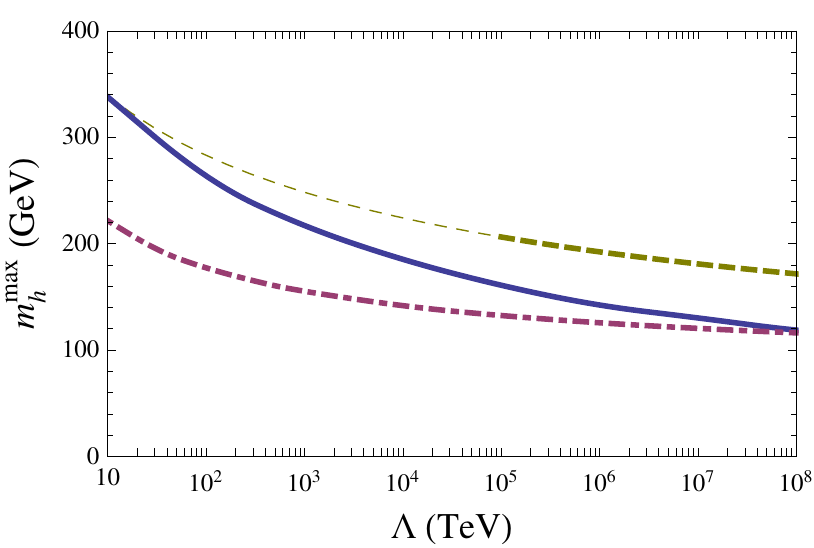}
\end{center}
\caption{{\small  Upper bounds on $m_h$ as function of the scale $\Lambda$ where some coupling starts becoming semi-pertubative, $g_x^2, g_I^2, \lambda^2 = 4\pi$ for the $U(1)$ case (dotdashed, $\tan{\beta} >> 1$), $\lambda$SUSY (solid,  $\tan{\beta} =1$) and $SU(2)$ (dashed, $\tan{\beta} >> 1$). In the $SU(2)$ case values of $m_h \gtrsim 270$ GeV are hardly compatible with naturalness and  the EWPT \cite{Lodone:2010kt}.}}
\label{MaxMhThreeModels}
\end{figure}

There is a price to pay, however, which explains the general consideration of the NMSSM only for $\lambda \lesssim 0.7$, i.e. not  $\lambda$SUSY\footnote{See \cite{Ellwanger:2009dp} and references therein}. A coupling $\lambda$ bigger than about 0.7 at the weak scale starts blowing up at a scale $\Lambda$ below unification, unless some change of regime occurs before.  Fig. 2 shows the  upper bound on the Higgs boson mass  that corresponds to $\lambda^2(\Lambda)$ being less than $4\pi$ \cite{Barbieri:2010pd}. Canonical gauge coupling unification is therefore under threat. The question then arises: Is this an unbearable step backward?  I do not think so, even apart from the relevance of the issues that $\lambda$SUSY addresses and potentially solves. It is conceivable that the three gauge couplings cross the threshold of a blowing interaction or even of a change in the relevant degrees of freedom without a significant change of their values, like it happens in the case of the fine structure constant going across the QCD scale. Furthermore I give weight to the fact that at $10^2\div 10^3$ TeV, a scale at which $\lambda$ could still  be semiperturbative with a Higgs boson above 200 GeV, the three gauge couplings are already all well within a factor of 2 from each other: $g_1 \approx 0.5, g_2 \approx 0.65$ and $g_3 \approx 0.9$, achieving a kind of {\it approximate} uniification. {\it Precise} unification is definitely appealing, but it may be not the right story.

\section{Addressing the flavour and the CP problems}

In view of  the current constraints,  the idea put forward long ago \cite{Dine:1993np,Pouliot:1993zm,Pomarol:1995xc,
Barbieri:1995uv,Cohen:1996vb,Dimopoulos:1995mi}
that the heaviness of the first two generations of sfermions may explain the lack of signals
from the plethora of potential new sources of flavour and CP violation present in the general MSSM is, {\it per se}, largely insufficient. If anything, some extra symmetry, compatible with the hierarchy in the sfermion mass spectrum, has to be operative to explain the supersymmetric flavour and CP problems. 

In \cite{Barbieri:2010ar} we propose that:
\begin{itemize}
\item Only the sfermions that interact with the Higgs system via the top Yukawa coupling are significantly lighter than the others.
\item With only the up-Yukawa couplings, $Y_u$, turned on, but not the down-Yukawa couplings, $Y_d=0$, there is no flavour transition between the different families, as in the SM case.
\end{itemize}
This follows in particular from assuming for $Y_u\neq 0$ but $Y_d=0$ that the quark sector has a flavour symmetry
\begin{equation}
U(1)_{\tilde{B}_1}\times U(1)_{\tilde{B}_2}\times U(1)_{\tilde{B}_3}\times U(3)_{d_R},
\label{U3}
\end{equation}
where $\tilde{B}_i$ acts as baryon number but only on the supermultiplets $\hat{Q}_i$ and $\hat{u}_{R_i}$ of the i-th generation, whereas $U(3)_{d_R}$ acts on the three right-handed supermultiplets of charge 1/3.
If one further assumes that, when $Y_d$ is switched on, it behaves as a non-dynamical spurion field that leaves the standard Yukawa couplings invariant under (\ref{U3}), it is easy to see that the  CKM matrix $V$ controls every flavour changing transition, in particular in the gaugino interactions with matter. I do not know of any other symmetry pattern that achieves this, while still be consistent with light stops and left-handed sbottom  only.

In this way all flavour and CP violating effects induced by loops of s-particle exchanges are under control.
 A  precise examination shows that the main contraints arise from  a mixed light-heavy exchange contribution to the $\epsilon_K$ parameter, controlled by the standard CKM phase, and from the one loop corrections to the electron and the neutron Electric Dipole Moments in presence of sizable flavour blind phases \cite{Barbieri:2011vn}. They require that the sfermions not coupled to the top be above $5\div 10$ TeV, depending on the values of the other parameters. In turn such values are consistent with a low fine tuning in $\lambda$SUSY, while they would not be in the MSSM\cite{Barbieri:2010pd}.
 
 With these values for the heavy sfermion masses, one remains with effective Minimal Flavour Violation and a moderate value of $\tan{\beta}$, as required by (\ref{m_h}) to have a heavy enough $m_h$.
The important residual signals are therefore related to the existence of sizable flavour blind phases and are expected to show up in EDMs at two loops or in CP asymmetries in B-physics, all mediated by the exchanges of the relatively light  stops and  left-handed sbottom \cite{Barbieri:2011vn}. It is noticeable that the supersymmetric flavour blind phases are precisely the sources of the unsolved CP problem in standard Minimal Flavour Violation based on $U(3)_Q\times U(3)_{u_R}\times U(3)_{d_R}$, i.e. with quasi degenerate squarks.

\section{Not the standard MSSM signals }

The picture outlined above and essentially represented in Fig. 2 has clear and distinctive phenomenological implications. As to the direct signals of supersymmetry, gluino pair production and subsequent decays into exclusive third generation quarks is the dominant feature. These final states may either result from chain decays with intermediate stops/sbottoms or from direct 3-body decays of the gluino. They are simply characterized by the gluino and the LSP mass, $m_{\tilde{g}}$ and $m_\chi$, and by effective Branching Ratios for the final states
\begin{equation}
\tilde{g}\rightarrow t\bar{t} \chi,~
\tilde{g}\rightarrow b\bar{b} \chi,~
\tilde{g}\rightarrow t\bar{b} \chi^- (\bar{t} b \chi^+),
\end{equation}
adding up to unity, with the last decay also followed by $\chi^\pm$ decays into $\chi$ plus a real or virtual $W^\pm$, depending on phase space\cite{Barbieri:2009ev}. With 5 $fb^{-1}$ of integrated luminosity the LHC at 7 TeV has a significant potential for discovery of these signals with gluino masses up to 1 TeV, at least in the most favourable cases for the other parameters \cite{Kane:2011zd}. 
As mentioned above, however, the extended naturalness region of $\lambda$SUSY, with gluinos above 1 TeV, may hide the direct supersymmetric signals to the LHC in its first phase.

This brings the focus on the search for the  Higgs boson(s), with in mind the projected discovery potential of the SM Higgs boson at the LHC in the next year or so. Discovering a Higgs boson with mass above 200 GeV would be in contradiction with the indication of the EWPT in the SM and definitely  incompatible with the MSSM. In turn, that this be possible or even necessary in $\lambda$SUSY depends on the gluon-gluon-$h$ coupling and on the $h$ decay branching ratios, with $h$ the lightest scalar. The studies so far show that the g-g-$h$ coupling, hence the production cross, is very close to the SM one for the same Higgs boson mass \cite{Cavicchia:2007dp,Franceschini:2010qz}. On the contrary the decay branching ratios are dependent on the composition of the $h$ boson, which may acquire a significant $SU(2)$-singlet component. It is in particular possible that the extra singlet pseudoscalar, $a$, be light enough, say around 100 GeV, to permit the decay $h\rightarrow a a$, with $a$ in turn decaying to a $b\bar{b}$ or $\tau\bar{\tau}$ pair \cite{Franceschini:2010qz}. A thorough analysis of the $h$ branching ratios in the general parameter space of $\lambda$SUSY may be worthwhile.

Finally it is worth to mention the impact of $\lambda$SUSY on the interpretation of Dark Matter in terms of neutralino LSPs and the related searches. Other than the fact that the neutralino LSP can have a significant singlet component, a Higgs boson with a mass above 200 GeV produces significant differences with respect to the MSSM picture. The first obvious one is the reduction of the neutralino-nucleous cross section, inversely proportional to the forth power of $m_h$, of relevance to the direct searches underway. The other has to do with  the relic  abundance of thermally produced neutralinos in the early universe: the $s$-channel exchange of a $h$ heavier than about 200 GeV in the neutralino annihilation cross section greatly influences the relic abundance of LSPs with mass in the 100 GeV range \cite{Barbieri:2010pd}.

In conclusion, while the MSSM remains a crucial benchmark for the search of supersymmmetry signals, it is important to keep in mind possible alternatives like the ones discussed above.

\section*{Acknowledgements}

I thank  Lawrence Hall, Slava Rychkov, Yasunori Nomura, Roberto Franceschini, Duccio Pappadopulo, Enrico Bertuzzo, Paolo Lodone, Marco Farina,  David Straub, Gino Isidori and Andrea Romanino for many useful discussions. This work was supported by the EU ITN ``Unification in the LHC Era'', contract PITN-GA-2009-237920 (UNILHC) and by 
MIUR under contract 2008XM9HLM. 

\bibliographystyle{utphys}
\bibliography{draft1}

\end{document}